\documentclass{elsart5}

\usepackage{graphicx}

\usepackage{amssymb}

\begin{document}

\begin{frontmatter}
% Title, authors and addresses
% use the thanksref command within \title, \author or \address for footnotes;
% use the corauthref command within \author for corresponding author footnotes;
% use the ead command for the email address,
% and the form \ead[url] for the home page:

\title{Electronic structure and transport properties of atomic NiO spinvalves} %%\thanksref{tit1}}
%% \thanks[tit1]{Title footnote}
\author{David Jacob\corauthref{cor1}} %%\thanksref{label1}}
\ead{david.jacob@ua.es}
%% \ead[url]{home page}
%% \thanks[label1]{footnote}
\corauth[cor1]{}
\author{J. Fern\'andez-Rossier}
\author{J. J. Palacios}
\address{Depto. de F\'isica Aplicada and Insto. Universitario de Materiales de Alicante (IUMA),
Universidad de Alicante, 03690 San Vicente del Raspeig, SPAIN} %% \thanksref{label2}}
%% \address[aff2]{Address}
%% \thanks[label2]{aff footnote}

\received{24 May 2006}
\revised{19 July 2006}
%% \accepted{14 June 2006}

%use optional labels to link authors explicitly to addresses:
%\author{}
%\address{}

\begin{abstract}
Ab-initio quantum transport calculations show that short NiO chains suspended 
in Ni nanocontacts present a very strong spin-polarization of the conductance.
The generalized gradient approximation we use here predicts a similiar polarization
of the conductance as the one previously 
computed with non-local exchange, confirming the robustness of the result.
Their use as nanoscopic spinvalves is proposed. 
\end{abstract}

%%%%%%%%%use  the \KEY command at the begin of keyword text%%%%%%%%%
%Please supply one or more relevant PACS-1996 classification codes 
%(http://publish.aps.org/PACS/96pacs.htmland) and about 5 keywords 
%of your own choice for indexing purposes. 
%You can see a list of already used keywords for JMMM at 
%http://authors.elsevier.com/JournalDetail.html?PubID=505704&Precis=KIND

\end{frontmatter}

%%\section{Introduction}
{\it Introduction-} 
While bulk nickel monoxide (NiO) is a common  example of a correlated insulator 
with antiferromagnetic (AF) order (see e.g. \cite{Sawatzky:prl:84,Moreira:prb:02}), 
short one-dimensional NiO chains can actually become half-metallic conductors, as 
we have shown recently with ab-initio quantum transport calculations \cite{Jacob:06}.
There we employed the hybrid density functional B3LYP as it gives a reasonable
description of the electronic and magnetic structure of bulk NiO \cite{Moreira:prb:02}
in contrary to the local density approximation (LDA) and the generalized 
gradient approximation (GGA) to density functional theory (DFT) \cite{Leung:prb:91}. 
B3LYP corrects the self-interaction error inherent in LDA and GGA by combining 
a (local) GGA exchange functional with the (non-local) Hartree-Fock (HF) exchange 
\cite{Becke:jcp:93}.

On the other hand, it is well known that metallic systems are better described
with a pure GGA functional. Thus it is not clear a priori which kind of functional will
give a better description of the composite system consisting of the atomic NiO 
chain and the metallic electrodes of a Ni nanocontact. 
It is therefore worthwhile to investigate the robustness of the obtained results
\cite{Jacob:06} for the NiO chains when a pure GGA functional is used.
Comparing the GGA results and B3LYP results further allows us to gain some insight 
on the effect and relevance of HF exchange in the B3LYP functional on the transport 
properties of the NiO chains.

%%\section{Method}
{\it Method-} 
The electronic structure of the infinite chain is calculated with the CRYSTAL
ab-initio program for crystalline structures \cite{CRYSTAL} while the electronic 
structure and transport properties of the Ni-O-Ni nanobridge in the Ni nanocontact 
is calculated with our ALACANT ab-initio quantum transport program\cite{Jacob:06}. 
The basis sets employed here are those of our previous work\cite{Jacob:06}. 
The GGA functional employed in the calculations is the one by Perdew and Wang\cite{PW91}.

\begin{figure}[t] 
  \begin{tabular}{cc}
    \includegraphics[scale=0.7]{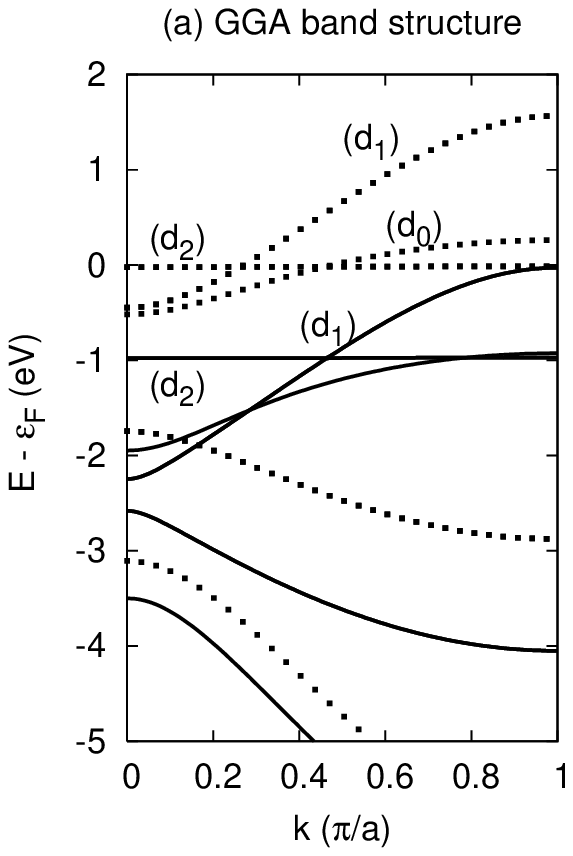} &
    \includegraphics[scale=0.7]{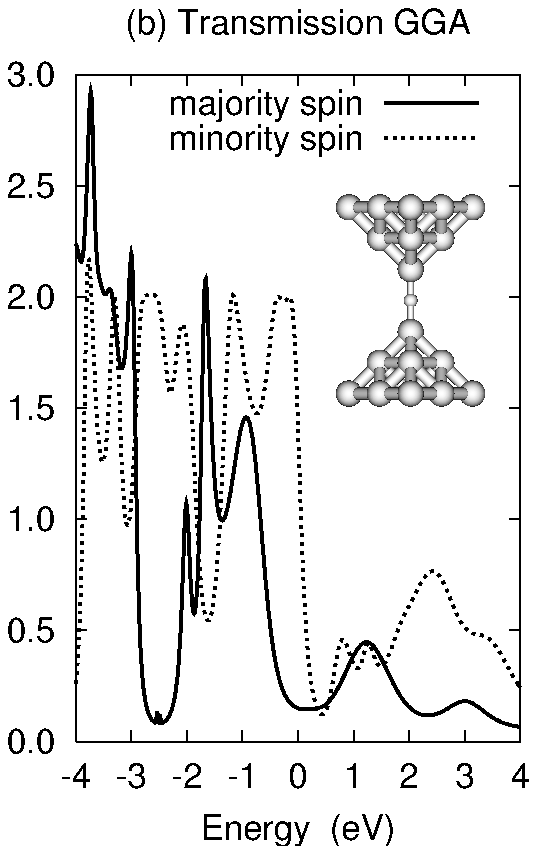}
  \end{tabular}
  \caption{
    (a) Spin-resolved band structure of one-dimensional Ni-O chain 
    in the FM phase. (b) Transmission per spin
    channel for Ni-O-Ni bridge in a Ni nanocontact consisting of two 
    pyramids in the 001 direction with 14 atoms each as shown in inset
    for parallel alignment of the electrodes' magnetizations.
    Solid lines indicate M bands/channels, dashed lines 
    m bands/channels. 
    Lattice parameter for the infinite chain and distance between
    Ni atoms in the Ni-O-Ni bridge are both 3.6\r{A}.
  }
  \label{fig-1}
\end{figure}

%%\section{Results and discussion}
{\it Results and discussion-} 
Within the GGA the one-dimensional NiO chain is always conducting 
for FM order in contrast to the B3LYP results where the FM state
can be either insulating or half-metallic\cite{Jacob:06}.
In Fig. \ref{fig-1}(a), we show the GGA band structure of an ideal 
one-dimensional NiO chain in the ferromagnetic (FM) phase. 
Compared to the B3LYP band structure for the half-metallic state 
at same lattic spacing \cite{Jacob:06} we see that the occupied 
bands have been raised considerably in energy. 
In particular, the doubly-degenerate flat minority-spin (m) band of type ($d_2$) 
composed of Ni $3d_{xy}$ and $3d_{x^2-y^2}$ orbitals (well below 
the Fermi level with B3LYP), now actually crosses the Fermi level. 
Also the doubly-degenerate and previously half-filled m band of type 
($d_1$) composed of Ni $3d_{xz}$ and $3d_{yz}$ orbitals hybridized with O 
$2p_x$ and $2p_y$ orbital (the only conduction band with B3LYP)
calculation are raised somewhat in energy.
Consequently, the previously empty m band of type ($d_0$) 
composed of Ni 3$d_{3z^2-r^2}$ orbitals is lowered in energy with 
respect to the other filled or partially filled $3d$ bands, and becomes 
a conduction band.

On the other hand also the majority spin (M) bands are raised in energy. The
doubly-degenerate M band composed of Ni $3d_{xz}$ and $3d_{yz}$
orbitals hybridized with O $2p_x$ and $2p_y$ orbitals which was well 
below the Fermi level in the B3LYP calculation now also crosses the Fermi level
near to its upper band edge.
Thus in GGA the ideal case of the infinite NiO chain in the FM phase 
does not represent a half-metallic conductor, although the spin-polarization
of the conduction bands is quite strong (5 m bands vs. 
2 M bands).

The change in the electronic structure of the one-dimensional NiO chain on the GGA level 
with respect to the B3LYP results can be explained by the insufficient cancellation 
of the self-interaction by the GGA exchange functional, which causes the occupied 
Ni $3d$ orbitals to artificially rise in energy. 

%%%%%%%%%%%%%%%%%%%%%%%%%%%%%%%%%%%%%%%%%%%%%%%%%%%%%%%%%%%%%%%%%%%%%%%%%%%%%%

%%%%%%%%%%%%%%%%%%%%%%%%%%%%%%%%%%%%%%%%%%%%%%%%%%%%%%%%%%%%%%%%%%%%%%%%%%%%%%

Finally, we have also calculated the electronic structure and transport properties 
of the single-oxygen atom bridge suspended between the tip atoms of a Ni nanocontact
making up a Ni-O-Ni nanobridge.
Fig. \ref{fig-1}(b) shows the transmission per spin-channel calculated with GGA for
the Ni-O-Ni nanobridge shown in the inset of Fig. 
We have checked that the results are stable with respect to the size of the
pyramidal Ni tips intervening the Ni-O-Ni nanobridge and the semi-infinite Bethe lattice
electrodes.
Surprisingly, the transmission does 
not look very different from the transmission calculated with B3LYP for that case. 
Near the Fermi level the transmission is strongly spin-polarized: The transmission
of the M channel is strongly suppressed while in the m
channel essentially two perfectly transmitting channels contribute to the conductance.

This seems to be at odds with the band-structure calculated for the infinite
one-dimensional chain, which suggests that there should be 5 m- and
2 M channels contributing to the overall conductance. 
However, the geometry of the Ni nanocontact blocks the transmission of just 
these channels which due to the unphyscial self-interaction of GGA have been
raised to the Fermi level.
Indeed, an orbital eigenchannel analysis \cite{Jacob:prb:06} of the transmission 
reveals that only the doubly-degenerate band of type ($d_1$) contributes to the 
conductance of the m channel just as in the case of the B3LYP functional.
The electrons in the flat m band of type ($d_2$) which actually crosses 
the Fermi level in the ideal case of the infinite chain are easily scattered as 
they present strongly localized electrons, and thus do not contribute to the 
overall conductance. 
The other m channel composed of Ni $3d_{3z^2-r^3}$ orbitals does not
contribute either to the conductance since the symmetry of the orbital is not 
compatible with the geometry of the two Ni electrodes - a mechanism to which we have 
referred to in previous work as {\it orbital blocking} \cite{Jacob:prb:05}.
The small but finite conductance in the M channel relates to the 
doubly-degenerate M band of type ($d_1$) of the infinite NiO chain 
raised to the Fermi level due to the self-interaction error. 
The ($d_1$) band only crosses the Fermi energy near the upper band edge where
the band becomes flat. Thus the the electrons in this channel are quite susceptible 
to scattering near the Fermi level resulting in a low transmission.

%%\section{Conclusions}
{\it Conclusions-} In summary, we have performed GGA calculations of the electronic structure and 
transport properties of atomic NiO chains.
Due to the insufficient cancellation of the self interaction by the GGA exchange
functional the density of states at the Fermi level is considerably higher for GGA
than for B3LYP and consequently the GGA conductance of the ideal infinite
NiO chain is much higher than the B3LYP conductance. This points to the importance
of correcting the self-interaction inherent in GGA for nanotransport calculations.
However, in the more realistic case of a short chain suspended in a nanocontact just 
those channels which have been raised to the Fermi level by the artificial self-interaction
are blocked by the scattering at the contacts. 
Due to this coincidence the GGA conductance is overall quite similiar to that of B3LYP.

% The Appendices part is started with the command \appendix;
% appendix sections are then done as normal sections
%% \appendix 
%% \section{}

\end{document}